# A Problem-Solving Framework to Assist Students and Teachers in STEM Courses

Jeffrey A. Phillips, Jeremy E. McCallum,

Katharine W. Clemmer, Thomas M. Zachariah

Loyola Marymount University

**Abstract**

Well-developed problem-solving skills are essential for any student enrolled in a science, technology, engineering or mathematics (STEM) course as well as for graduates in the workforce. One of the most essential skills is the ability to monitor one's own progress and understanding while solving a problem. Successful monitoring during the entire solution allows a solver to identify errors within a solution and make adjustments as necessary. To highlight this aspect of problem-solving, we have developed a framework and associated classroom activities that introduce students to monitoring (M) alongside the more traditional aspects of problem-solving models: analyzing the task (A), creating a plan (C), and executing the plan (E). This ACE-M framework has been successfully implemented in lower-division chemistry, mathematics and physics courses. Students enrolled in courses where ACE-M was used as the foundation for problem-solving instruction reported improved problem-solving self-efficacy, more monitoring while solving problems, and in many cases improved course grades. With this explicit instruction on self-monitoring, students are now introduced to expert problem-solving skills that will benefit them in their STEM careers.

# Introduction

Problem-solving involves reaching a goal despite the solution pathway being unknown at the outset. (Newall & Simon, 1972; Bransford & Stein, 1993) Because of this inherent uncertainty, a solver cannot simply follow a known solution pathway and must improvise throughout the process. In order to reach the desired goal, a solver should continuously and objectively examine his/her actions. This monitoring and evaluating allows the solver to regulate and error check his/her thought process. Such monitoring or control is essential to refining the solution while crafting it. (Son and Schwartz, 2002)

In the workplace, employees often are asked to solve problems in the midst of changing conditions and goals. Because of this need for independent thinking, problem-solving skills are seen as essential learning outcomes for all college graduates. (AAC&U, 2007) Many employers consider the quality of a person's critical thinking skills, such as problem-solving, a more important factor in the hiring process than a candidate's major. (Hart, 2013)

While employers rate problem solving/critical thinking as one of the top five "very important" skills for job success, only 28% classify college graduates' problem solving as excellent. (Casner-Lotto & Barrington, 2006) Another recent survey showed a similar perspective, with 400 employers saying that only 24% of the recent college graduates were well-prepared to engage in analyzing and solving complex problems. (Hart Research Associates, 2015) Other surveys similarly describe the need for improved problem-solving instruction, reporting that 64% and 82% of employers desire a greater emphasis on complex problem-solving in college. (AAC&U, 2007; Hart, 2013)

Many teachers and scholars have attempted to respond to this need for improved problem-solving skills among college graduates over the years. One of the earliest, and most influential, scholars to do this is George Pólya. In his classic text, *How to Solve It*, Pólya (1957) espoused a four-step problem-solving process: 1. Understand the problem, 2. Make a plan, 3. Carry out the plan, and 4. Look back on your work. While Pólya did recommend some reflection at the end to help the solver understand what worked and what didn't, his suggested process does not emphasize the necessary monitoring that must occur throughout the process in order to successfully create solve a problem. In physics, Reif et al. (1976) developed a similar four-step process to guide students.

To strengthen students' systematic approach, particularly in the early portion of a solution, many problem-solving frameworks expand the number of steps. For example, Heller and Heller (1995) include the steps "Focus the Problem" and "Describe the Physics" before "Plan a Solution." Bunce et al. (1991) include distinct steps that ask students to identify the givens, the goal and relevant principles in their six-step Explicit Method of Problem Solving (EMPS) process. While these more detailed processes increase the probability that a student will mimic the expert behavior of categorizing a problem before articulating a plan, they omit the key monitoring process.

Most introductory science, technology, engineering, and mathematics (STEM) textbooks describe problem-solving as a four- or five-step process. (Phillips, et al., 2015) While the labels vary, they are very similar to the above processes. Nearly all imply that problem-solving is a linear rather than recursive process. While several of the above processes ask the students to check the final numerical answer and/or reflect at the end, none clearly call out the need for the solver to monitor his/her thoughts throughout the solution pathway.

Bransford and Stein developed the IDEAL method of problem-solving, which includes the step "Explore Alternative Approaches." (Bransford & Stein, 1993) While this does encourage students to do some monitoring, it does not strongly encourage different ways of monitoring throughout the solution process. (Phillips, Osorno & Fier, 2013) There are other models of problem-solving that include monitoring and other components such as confidence and creativity, (Adams and Wieman, 2015) but these are likely too complex for teachers and students to use as a tool in the classroom.

**ACE-M Framework**

When a solver does not know a solution pathway immediately, he/she must employ monitoring to determine if a problem has been correctly analyzed or if a plan is likely to be successful. This monitoring is important in problem solving and all aspects of creating knowledge. (Schoenfeld, 1985; Schoenfeld, 1987) Problem solvers who have a greater self-awareness and act intentionally succeed more often than those who are not reflective and approach the problem in a piecemeal fashion. (Bransford, Brown & Cocking, 1999) Without using monitoring, students often use the first idea that occurs to them, whether or not it is appropriate, and fail to explore alternatives. (Larkin, et al., 1980; Mestre, et al., 1993) Most students do not receive sufficient instruction on how to develop these skills as instructors and texts often emphasize strategic and procedural knowledge, even though the major difficulties lie with monitoring and control. (Mayer, 2008)

There are two aspects to successful monitoring- "self-monitoring" and "task-monitoring." When self-monitoring, the solver reflects on what he/she understands, feels uncomfortable with, and is capable of doing. This awareness, often referred to as metacognition, can lead to the solver to change pathways to avoid a technique that is too challenging, or realize that a new technique or concept has been mastered. (Flavell, 1971) When task-monitoring, the solver is continuously examining the solution in order to regulate and control the process. This critical skepticism is similar to what an objective study partner might do. This second form of monitoring usually involves asking questions such as: "do ideas X and Y agree with each other?"; "does this calculation agree with what the question provided?"; or "is X a more appropriate model to use than Y?"

To emphasize the importance of monitoring to students, we have placed it within our framework on equal footing with the various cognitive aspects of problem-solving. Our framework, *ACE-M*, has four components: *Analyze the Task*, *Create a Plan*, *Execute the Plan*, and *Monitor Understanding and Actions*. It is important to note that Monitoring is not a fourth, separate component; rather, it is integrated into the other components. To more clearly illustrate what monitoring entails, examples are included in the A, C and E descriptions found below:

> **A**: *Analyze the Task–* Interpret and understand what is provided in the task/problem. Suggestions often include annotating the task, creating alternate representations, and identifying the relevant information given in the task, including the task goal.

- Monitoring: Ask if any of the given information is irrelevant or contradictory, consider alternative ways of interpreting the task and/or goal, or contrast the given task with previously seen ones. Ask where points of confusion remain.

**C**: *Create a Plan–* Connect the given information and desired result with models/concepts/relationships. Solvers make assumptions and establish connections and intermediate steps between the given information and goal.

- Monitoring: Compare the plan with others used in previous tasks, ask if any of the assumptions or models are incompatible with the given information, or consider alternative plans.

**E**: *Execute a Plan–* Follow the plan until the desired result is attained.

- Monitoring: While executing, continuously examine the plan to ensure that it is internally coherent, ponder alternative ways of executing the plan or ask if the task has been completed. At the end of the process, considering if your solution pathway will work in all cases.

**M**: *Monitoring Understanding and Actions–* Ask questions to determine when there is an error, when to proceed to a new phase of the solution pathway (task-monitoring), or when there is something missing or misunderstood in one's own knowledge, skill or belief structure (self-monitoring). The inclusion of monitoring transforms what might be a monologue into something resembling a dialogue as the solver constantly seeks to identify ways to improve the solution and his/her understanding through the verbalization of questions and decisions.

Despite what may be implied by the acronym, instructors stress to students that the ACE-M framework implies a non-linear or recursive process. Since monitoring is occurring throughout the process, a solver may discover the need to engage in a different component of the process before moving forward. Also, simply altering the language from problem-solving "steps" to "components" helps many students avoid incorrectly viewing problem-solving as a linear process that is without backtracking.

**Classroom Implementation**

One of the ACE-M framework's primary functions is scaffolding students' problem-solving. In this way the framework becomes a tool that aids students. Throughout a course, students are introduced to, and generate, prompts that aid them within each component. For example, "When you were first analyzing the question, did you ask if any of the given information is contradictory?"; "Do various assumptions agree with each other and the interpretation of the task?"; "Are there other possible plans that should be utilized?"; and "When thinking through your execution, did you compare the result of the plan against previous experiences?" Students learn to use the prompts on all problems as instructors and classmates will often ask for the answers in discussions.

The ACE-M framework's usefulness can extend to exams where students are asked to separate out their thoughts into the appropriate component. By associating each component with points, students can see the value the instructor places on the entire process. Typically the *Analyze the Task* and *Create a Plan* components are worth 40% each, which is often different from students' initial view that emphasizes calculations in the *Executing a Plan* component. While it doesn't factor into the required components,

*Monitoring Understanding and Actions* also is part of the exam structure. Students can write their questions and observations for partial credit under any of the three required components.

Instructors model problem-solving by utilizing a think-aloud protocol. They not only describe what actions they are taking, but also describe alternative pathways. Instructors will spend as much time describing the pros and cons of pathways that were not taken as those that were. Often solutions start down unproductive pathways, but through the monitoring of their actions, instructors detect the errors and illustrate how to improve the solution. This monitoring-rich modeling occurs in nearly every worked example in class. To augment the monitoring-poor worked examples of a textbook, many instructors also record think-aloud solutions. Using smartpens or tablets, it is possible to capture what is written on a page (or screen) and what is vocalized by the solver. (http://www.livescribe.com and https://www.educreations.com) These movies more clearly illustrate the problem-solving process than polished solutions.

To facilitate feedback on their problem-solving, students are also asked to solve problems using a think-aloud protocol. This includes some, if not all, of the homework assignments. Students use the same technology as the instructors to record think-alouds out of class. Seeing the problem-solving process, including errors and insights, lets the instructor provide meaningful feedback, which cannot be generated from the end product- a static, sanitized solution. By shifting the in-class activities and homework to think-aloud protocol, instructors are able to reinforce the message that what is important is the process, rather than the "final answer."

Most students are not particularly proficient at performing think-aloud solutions at the beginning of a course. To address this, instructors devote class time and assignments to practicing think-alouds. The feedback in the initial activities is solely about the quality of the solver's verbalization, not the content. Early assignments may be grounded in non-STEM content that is more comfortable for the students, e.g. "teach the listener how to___," or "icebreakers" where each student contributes and justifies his/her opinions, such as selecting the characters who are allowed to escape a hypothetical catastrophe.

The emphasis on think-alouds stems not only from the desire to provide feedback to students, but also a belief that expressing one's thoughts leads to improving the monitoring of them. This is because articulating one's thoughts forces one to evaluate and refine them, aka monitor them. Solvers who articulate their decision-making as they solve problems, are able to solve them more effectively than those who don't. (Ahlum-Heath and DiVesta, 1986; Beradi-Coletta, et al., 1995) Simply talking out one's actions is not as effective. The focus of the verbalization must be on the decision-making, that is, the questions and answers that comprise monitoring. By articulating one's thinking, monitoring occurs more frequently and productively and the solution is refined. Think-aloud solutions are a tool by which instructors scaffold students' monitoring.

When students work together in class on solutions, one student is often assigned the role of solver and the other(s) assigned the role of listener(s) or skeptic(s). Those in listener role are instructed to prod the solver to verbalize his/her thoughts if he/ she is silent for more than 5 seconds. As the course progresses, the second role evolves into that of a skeptic who asks more probing questions. These go beyond what would be asked in a think-aloud that is used for research. For example, skeptics may ask: "I don't quite understand why you did that."; "What are you trying to accomplish now?"; "How are you deciding what to do next?"; or "What other ways can you approach this solution?" The listeners and skeptics help the solver refine his/her ideas as they ask for clarification. (Johnson and Chung, 1999; Whimbey and Lochhead 1999) By the end of the course, the skeptic essentially becomes the voice of monitoring and students are engaged in dialogues that mimic what a person must do internally to be a successful problem-solver.

**Results**

The ACE-M framework has been successfully used in lower-division chemistry, mathematics, and physics courses. The majority of students have described changes in their problem-solving habits, which they attribute to ACE-M. There is also evidence that ACE-M and the associated activities have positively impacted students' performance.

By the end of the course, 85% of the Calculus 1 students shared that ACE-M was the most important process they learned to support their problem-solving efforts. When asked if the ACE-M process was helpful, 87% of General Physics students reported that it was. Below are some representative quotes from General Physics students:

- "The ACE-M method got me on track and helped me on difficult problems that I would've been lost on."
- "At the beginning of the semester, I was ok at solving problems, but most of the time I would just plunge ahead without really thinking about if what I was doing made sense. Now I try to stop myself every once in a while to check and make sure that the equations I'm using and numbers I am getting make sense."

- "I think my problem-solving abilities have definitely improved after this course. I am now able to look at a problem more critically and better understand what the question is really asking for."

To observe whether or not students in a section that utilized the ACE-M framework and activities viewed problem-solving differently than those in other sections, students in both completed pre-instruction and post-instruction surveys. Twelve Likert scale items assessed students' self-efficacy in solving mathematical and scientific problems. Students in ACE-M Calculus and Organic Chemistry courses reported a statistically significant improvement ($p< 0.05$) between their pre-instruction and post-instruction responses whereas those in comparison sections did not.

Students were asked to rate their use of monitoring when solving problems by responding to twelve Likert scale items. When comparing the responses of the post-instruction survey to those of the pre-instruction survey, it was seen that students in the ACE-M Calculus and Organic Chemistry courses reported gains in their value and use of monitoring that were statistically significant ($p< 0.05$), whereas those in comparison courses indicated no such gains.

ACE-M not only impacted students' self-efficacy and use of monitoring tactics, it also positively impacted their performance. On a post-instruction survey, students were asked "What aspects of this class and/or your teacher's instruction was most helpful to improve your ability to solve problems?" Calculus students who explicitly cited ACE-M (or even just one of the components) had course grades that were almost a full letter grade higher than those who didn't mention ACE-M. This statistically significant difference ($p< 0.05$) indicates that those who actually understand and value a methodical and complete problem-solving process have greater success in their class.

Students who were enrolled in a Calculus 1 course that was structured around the ACE-M framework were able to realize the benefits not only in that course, but also the next one, where they had different instructors. Students who had been taught with the ACE-M framework and activities had an average Calculus 2 grade of 3.09, which exceeded the 2.81 average earned by a comparable population taking the same class. The techniques that the students learned transferred to a different course and helped them succeed. Physics students who were taught ACE-M also saw a similar benefit over their peers in subsequent courses, but the effect was smaller and not statistically significant.

**Conclusion**

Solving problems is a learning goal that is central to all STEM disciplines, yet instruction on accomplishing this goal is often incomplete. Central to the problem-solving process is the ability to self-correct the inevitable errors. A solution free of errors likely means that the solver was not presented with a true problem– a task where the solution pathway is not known to the solver at the start. Without successful monitoring, one is incapable of detecting errors or making adjustments.

Teachers rarely instruct students on how to monitor their problem-solving. In response to this gap, we developed the ACE-M framework and associated activities to place the emphasis on monitoring. When solving problems, students are provided scaffolding in the form of prompts. Instructors model successful problem-solving with great attention paid to the decisions when solving problems. This modeling, and much of the coaching, relies on a think-aloud protocol where solvers articulate their thinking. This constant verbalization also serves as a tool that aids students. By learning how to articulate their thinking, students are able to refine and improve it.

In chemistry, mathematics, and physics courses, we have observed positive impacts on students who utilized ACE-M when solving problems. Many students reported improved problem-solving self-efficacy as well as a greater usage of monitoring. Other students demonstrated improved performance in their lower-division STEM courses, including those in which they enrolled later. By devoting class time to the problem-solving process, rather than the final polished product, students were able to observe and receive feedback on all of the skills that they need to succeed.

**Acknowledgment**

We are grateful to the National Science Foundation for award DUE 10-44062.